\documentclass[aps,prl,twocolumn,groupedaddress,amssymb,amsmath,amsthm]{revtex4-1}
\usepackage{amsfonts}

\usepackage{amssymb}

\usepackage{verbatim}
\usepackage{amsmath}
\usepackage{graphicx}

\begin{document}
\title{Bose Gases Near Unitarity}
\author{Weiran Li and Tin-Lun Ho}
\affiliation{Department of Physics, the Ohio State University,
Columbus, Ohio, 43210}
\date{\today}
\begin{abstract}
We study the properties of strongly interacting Bose gases at the density and temperature regime when the three-body recombination rate is substantially reduced. 
In this regime, one can have a Bose gas with all particles in  scattering states (i.e. the  ``upper branch") with little loss even at unitarity over the duration of the experiment. 
We show that because of bosonic enhancement, pair formation is shifted to the atomic side of the original resonance (where scattering length $a_s<0$), opposite to the fermionic case.  In a trap, a repulsive Bose gas remains mechanically stable when brought across resonance to the atomic side until it reaches a critical scattering length $a_{s}^{\ast}<0$. For $a_s<a_{s}^{\ast}$, the density consists of a core of upper branch bosons surrounded by an outer layer of equilibrium branch. The conditions of low three-body recombination require that the particle number $N< 1.024 (T/\omega)^{5/2}$ in a harmonic trap with frequency $\omega$. 
\end{abstract}
\maketitle

One of the most fascinating aspects of quantum gases is the role of strong interaction, which is generated by bringing the system close to a Feshbach resonance\cite{Chin}. 
The scattering between particles near resonance leads to a very large negative or positive s-wave scattering length $a_{s}$, causing strong attraction or repulsion between 
atoms in scattering states. In the case of fermions, preparing the system on the atomic side of the resonance ($a_{s}<0$), the ground state of the system exhibits BCS-BEC crossover as the system is brought across the resonance to the molecular side ($a_{s}>0$)\cite{Stringari}. 
In contrast, a repulsive Fermi gas is metastable, as fermions with positive scattering length can form bound states (Feshbach molecules) through  three-body recombination.  Once Feshbach molecules are formed, they can decay into deep bound states through collisions, which leads to atom loss.  

In the two few years, there have been active experimental studies on strongly repulsive fermions, driven by the quest  of Stoner ferromagnetism\cite{MIT}. While it is now shown that ferromagnetism is  absent in  repulsive fermions\cite{MIT2}, many earlier experiments performed over a wide range of physical conditions all show similar behavior in atom loss \cite{Shizhong, Pekker}. Moreover,  an early experiment\cite{Salomon2003} has clearly demonstrated that the energy density exhibits a maximum near resonance on the molecular side of the resonance, which is found to be caused by Pauli blocking on the formation of bound pairs\cite{vijay}. 

In the case of Bose gases, attractive interactions will cause mechanical instability at low temperatures. Thus, most studies focus on repulsive Bose gases. However, like repulsive Fermi gases, repulsive Bose gases are only metastable. For weak interactions, the collision rate due to three-body  recombination  ($\gamma_{3}= - n^{-1}{\rm d}n/{\rm d}t)$ is $\gamma_{3}= c (4\pi \hbar a_{s}/m)n (na_{s}^3)$, where $c$ is a dimensionless constant\cite{Petrov}, while the two-body collision rate is $\gamma_{2} = na_{s}^2 v$ where $v$ is the typical velocity of the bosons. 
For weak repulsion, $n^{1/3}a_{s}\ll 1$, $\gamma_3$ is sufficiently low that the system is essentially free of molecules.   In the last two years, there are increasing number of experiments on strongly repulsive Bose gases at low temperatures\cite{Hulet, Cornell, Salomon-boson}.  However, at low temperatures, $\gamma_{3}$ increases rapidly in the  strongly repulsive regime, i.e. $n^{1/3}a_s  >1$. This leads to severe atom loss as the system approaches resonance, and the system is far from equilibrium. While one can explore strong interaction effects by bringing the system quickly in and out of the strongly interacting region, it is not clear how to define equilibrium properties in such situations.

The situation is different at higher temperatures and lower densities, i.e.  lower fugacities. At temperatures $T>T_{c}$, we have  $v\sim \sqrt{3k_{B}T/m}\sim h/(m\lambda)$, where  $T_{c}$ is the  the BEC transition temperature, and  $\lambda=h/\sqrt{2\pi mk_{B}T}$ is the thermal wavelength. Close to unitarity, 
 $a_s$ in $\gamma_2$ and $\gamma_3$ is replaced by $\lambda$ in this temperature regime, and we have $\gamma_{2} = (k_{B}T/\hbar) (n\lambda^3)$, and  $\gamma_3 = {\cal C}  (k_{B}T/\hbar) (n\lambda^3)^2$, where ${\cal C}=9\sqrt{3}/\pi \sim 4.96$\cite{c}.   As density drops, $\gamma_3$ will eventually fall below $\gamma_2$. And in the presence of a trap, the spatially averaged rate of total particle loss, $-N^{-1}{\rm d}N/{\rm d}t = \int {\rm d}{\bf r} \gamma_{3} n/\int {\rm d}{\bf r} n = \langle \gamma_{3} \rangle_{ave}$, will fall below the trap frequency $\omega$, where $\langle .. \rangle_{ave}$ means spatial average. 

 In the density and temperature regime (referred to as ``low-recombination" regime) where $\gamma_3 < \gamma_2, \,\, \langle \gamma_{3}\rangle_{ave} <\omega$,  or 
 \begin{equation}
  n\lambda^3<<1, \,\, \,\,\,\,\,  \overline{n}\lambda^3 < {\cal C}^{-1/2} \sqrt{\hbar \omega/k_{B}T},
  \label{conditions}  \end{equation}
where $\overline{n}^2 \equiv \int n^3/\int n$, very few molecules are formed  {\em even at unitarity} during  the time when the Bose gas reaches global equilibrium through two-body collisions.  
 We can then reach an equilibrium state where the bosons are in scattering states even though the system can accommodate Feshbach molecules. 
This ``low-recombination" regime has recently been realized by Salomon's group at ENS\cite{Nir-private}. 
In this paper, we shall point out a number of surprising properties of strongly interacting Bose gases in this 
 low-recombination regime. We  find that

\noindent {\bf (I)} Bose  statistics enhances pair formation.
As a result, molecule formation in a homogenous Bose gas is shifted to the atomic side ($a_s<0$), in contrast to fermions where the shift is  to the molecular side ($a_s>0$) due to Pauli blocking\cite{vijay}.  The energy change of the system when making transition from the upper to lower branch (defined later) is substantial even at temperatures as high as $10T_c $.

\noindent {\bf (II)}  In a trap, when a repulsive Bose gas is brought across resonance in the low-recombination regime, it remains stable even on the atomic side $(a_s<0)$, but up to a  critical value $a_{s}^{\ast}<0$, and its density  consists of a metastable 
``upper branch" core surrounded by an outer layer of bosons in thermodynamic equilibrium.  Both regions are molecule free. The system will suffer mechanical instability 
for $a_s> a_{s}^{\ast}$. 

\noindent {\bf (III)}  The conditions for low-recombination at unitarity (Eq.(\ref{conditions})) and mechanical stability constrain the total number of particles in a trap. In order to observe 
the phenomena in ${\bf (II)}$, we need $N< \alpha N^{\ast}$,  $N^{\ast}= (k_{B}T/\hbar \omega)^{5/2}$ where the constant $\alpha=1.024$. For an estimate at $T=1 \mu$K, $\omega = 2\pi (250) {\rm sec}^{-1}$, we have $N^{\ast}\sim 6.5\times 10^4$. 

{\em (A) Homogenous upper branch Bose gas:} 
We first study the homogenous repulsive Bose gases that are free of molecules. 
Such system will be referred to as ``upper branch" Bose gas, and is a good approximation of a Bose gas in the low-recombination regime. In contrast, the equilibrium state of a Bose gas consisting of both atoms and molecules will be referred to as the ``lower branch" or ``equilibrium branch".  
To study the upper branch Bose gas,  
we use a generalized Nozieres Schmitt-Rink(NSR)\cite{NSR} method recently developed by one of us (TLH) for the upper branch Fermi gas\cite{vijay}. 
(We set  both  $\hbar$ and $k_{B}$ to 1 from now on).  It is straightforward to see that the equation of state is identical to that of a Fermi gas, except that all the Fermi functions are replaced by the Bose distribution functions $n_B(\omega)=1/(e^{\omega/T}-1)$. The result is $n(\mu, T)= n_{o}(\mu, T)+\Delta n^{sc}(T,\mu)  + \Delta n^{bd}(T,\mu)$, where $n_{o}(T,\mu) = \sum_{\bf k} n_B(\xi_{\bf k})$ is the density of the ideal Bose gas;  $\Delta n^{sc}(T,\mu) $ and  $\Delta n^{bd}(T,\mu)$ are the interaction contributions of the scattering states and the bound states respectively, 
\begin{equation}
\Delta n^{sc}(\mu,T)=  -\frac{1}{\Omega}\sum_\mathbf{q}\int^{\infty}_{\omega(q)} \frac{\mathrm{d}\omega}{\pi}\, n_B(\omega)\frac{ \partial \zeta ({\bf q}, \omega)}{\partial \mu},
\label{nsc} \end{equation}
\begin{equation} 
\Delta n^{bd}(\mu,T)= -\frac{1}{\Omega} \sum_\mathbf{q} n^{}_{B} (\omega_{b}(q))
\frac{ \partial \omega_{b}(q)}{\partial \mu}, 
\label{nbd} \end{equation}
where $\omega(q)\equiv q^2/4m -2\mu$.  $\zeta ({\bf q}, \omega)$ is the phase of the inverse 
 $T$-matrix in a medium whose explicit expression is given in ref.\cite{vijay}. It arises from the branch cut of the $T$-matrix (i.e. the scattering states).  
 $\omega_{b}(q)$ is the pole of the $T$-matrix (i.e. the bound states), and is the solution of the equation 
\begin{equation}
-\frac{m}{4\pi a_s} + \frac{1}{\Omega} \sum_{\mathbf{k}} \left(\frac{\gamma({\bf k}; {\bf q})}{\omega-\omega(q)-\frac{k^2}{m}}+\frac{1}{\frac{k^2}{m}}\right)=0,
\label{pole} \end{equation}
where
$\gamma({\bf k}; {\bf q})= 1+n_B(\xi_\mathbf{q/2+k})+n_B(\xi_\mathbf{q/2-k})$ describes the bosonic enhancement of the medium on pair fluctuations, $\xi_{k} = \epsilon_k-\mu$,  and $\epsilon_k=k^2/2m$.

As pointed out in ref.\cite{vijay},  the NSR results Eqns.(\ref{nsc})  and (\ref{nbd}) reduce to the  
scattering state and bound state contributions in the rigorous virial expansion in the low fugacity limit. 
The equation of state for the upper branch at lower temperatures therefore corresponds to ignoring $\Delta n^{bd}$, whereas that of the equilibrium state (the lower branch) includes both $\Delta n^{sc}$ and $\Delta n^{bd}$; 
\begin{eqnarray}
n_{upper}(T, \mu) &= &n_{o}(T, \mu) + \Delta n^{sc}(T, \mu) \hspace{0.3in}  \label{n-upper} \\
n_{equil} (T, \mu)   &=& n_{o}(T, \mu) + \Delta n^{sc}(T, \mu) +  \Delta n^{bd}(T, \mu).\hspace{0.2in} \label{n-lower}
\end{eqnarray}
With this prescription, we can calculate all the thermodynamic properties of these branches\cite{MuellerBaymStoof}. Our results are summarized in the next two sections.

\begin{figure}
\includegraphics[width=0.45\textwidth]{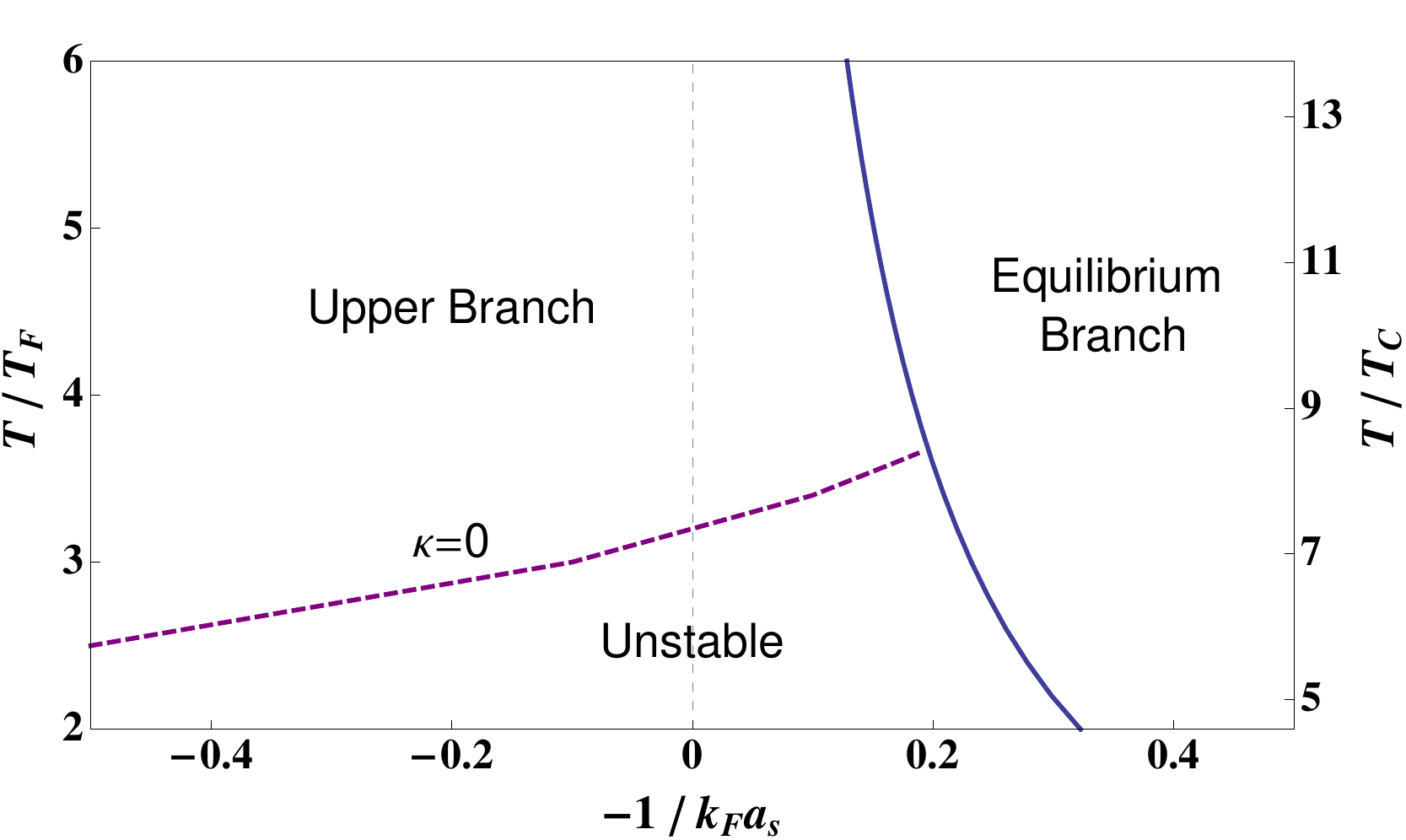}
\caption{\label{fig:homoPD} The phase diagram of a homogeneous upper branch Bose gas with fixed density $n$: At the blue curve that separates the upper and lower branch, the energy density undergoes a discontinuous jump as shown in Fig.\ref{fig:energy}. The purple dashed curve represents a state with $\kappa=0$. In the ``unstable" region, the number equation for chemical potential does not have a solution.}
\end{figure}

\begin{figure}
\includegraphics[width=0.35\textwidth]{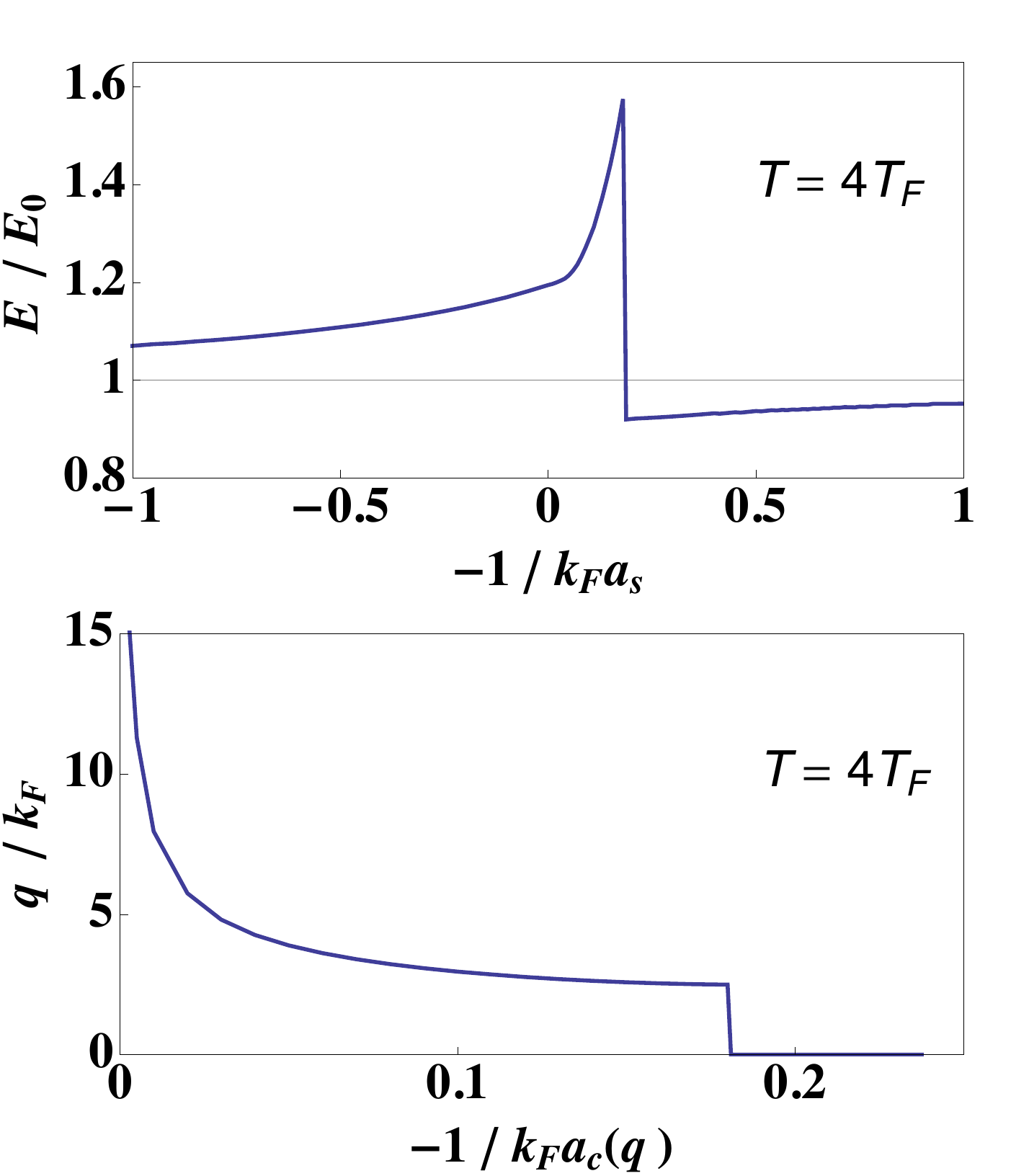}
\caption{\label{fig:energy}The upper panel is the energy density across resonance at $T=4E_F=9.2T_{c}$: rescaled by energy $E_{o}(T)$ of a noninteracting system at the same temperature. The jump represents a transition from the upper to the lower branch. 
Even at this high temperature, interaction energy and  the jump are substantial fractions of the total energy.  The lower panel shows $a_c(q)$ as a function of $q$ for fixed $n$. The jump is due to the sudden change in $\mu$ as the system switches branches.}
\end{figure}

{\em (B) Phase diagram of homogenous upper branch Bose gas:} It is useful to define for bosons an analog of  ``Fermi" momentum and ``Fermi" temperature as $k_{F} \equiv (6\pi^2n)^{1/3}$ and  $T_{F} \equiv k^{2}_{F}/2m$; and $T_{F}/T_{c}= 2.3$, where  $T_{c}=3.3 n^{2/3}/m$ is the BEC transition temperature. 
The phase diagram of the upper branch Bose gas for fixed $n$ is shown in Figure \ref{fig:homoPD}. The corresponding behavior of the energy density at $T=4T_{F}$ is shown in Figure \ref{fig:energy}. 

The dashed purple line in Fig. \ref{fig:homoPD} corresponds to the state of zero compressibility $\kappa =0$, where $\kappa= dn/d\mu$. While the decrease in $\kappa$ as $a_s>0$  increases is similar to that of Fermi gas, this phase diagram differs from that of Fermi gas in a fundamental way, as the transition from upper to lower branch is shifted to the atomic side of the resonance, ($a_s<0$). 
See Fig.\ref{fig:homoPD} and Fig.\ref{fig:energy}. In other words, the  stability of the upper branch Bose gas extends into the atomic side. 
This shift is due to Bose statistics.   To see this, we note that the condition for determining the emergence of a bound state in the medium (i.e. Eq.(\ref{pole})) can be rewritten as  
\begin{equation}
\frac{m}{4\pi a_{eff}({\bf q}, \omega)} = \frac{1}{\Omega} \sum_{\mathbf{k}} \left(\frac{1}{\omega-\omega(q)-\frac{k^2}{m}}+\frac{1}{\frac{k^2}{m}}\right), 
\label{equiv} \end{equation}
\begin{equation} \frac{1}{a_{eff}({\bf q}, \omega) } = \frac{1}{a_{s}}   - \frac{4\pi}{\Omega} \sum_{\mathbf{k}} \left(\frac{n_B(\xi_{{\bf q}/2+{\bf k}})+n_B(\xi_{{\bf q}/2-{\bf k}})}{m(\omega-\omega(q))-k^2}\right); 
\label{aeff}  \end{equation}
which is  the condition for emergence of bound state in vacuum with an effective scattering length $a_{eff}({\bf q},\omega)$. In vacuum, $a_{eff}$ reduces to $a_s$, and the bound state occurs when $a_{s}^{-1}\geq 0$. In a Bose medium, a bound pair with total momentum ${\bf q}$ occurs when $a_{eff}({\bf q}, \omega=\omega(q))\geq 0$, or  when $-a_{s}^{-1}\leq  -a^{-1}_{c}(q)$, where 
\begin{equation}
\frac{1}{a_{c}(q)}=  -
  \frac{4\pi}{\Omega} \sum_{\mathbf{k}} \frac{n_B(\xi_{{\bf q}/2+{\bf k}})+n_B(\xi_{{\bf q}/2-{\bf k}})}{k^2} <0.   
\label{as-q} \end{equation}
In other words, if one approaches the resonance from the atomic side, a bound pair with total momentum $q$ will emerge at scattering length $a_{c}(q)<0$, which is on the atomic side of the original resonance. 
 The values of $-1/a_{c}(q)$ at different $q$'s are shown in Fig.\ref{fig:energy}.  That $-1/a_c(q)$ reduces to  0 as $q\rightarrow \infty$ is because the effect from the bosonic medium becomes less important for large $q$, as in fermion case\cite{vijay}.
 The boundary between the equilibrium branch and the upper branch 
 corresponds to the emergence of bound pairs with $q=0$, and is given by $a^{-1}_{c}(q=0,T, \mu)=0$, where $\mu$ is constrained by the total density $n$.  
Finally, we note from Fig.\ref{fig:energy} that as one crosses the resonance from the molecular to the atomic side, the energy change at the boundary between upper and lower branch is substantial even at $T=4T_F= 9.2T_c$. 

\begin{figure}
\includegraphics[width=0.45\textwidth]{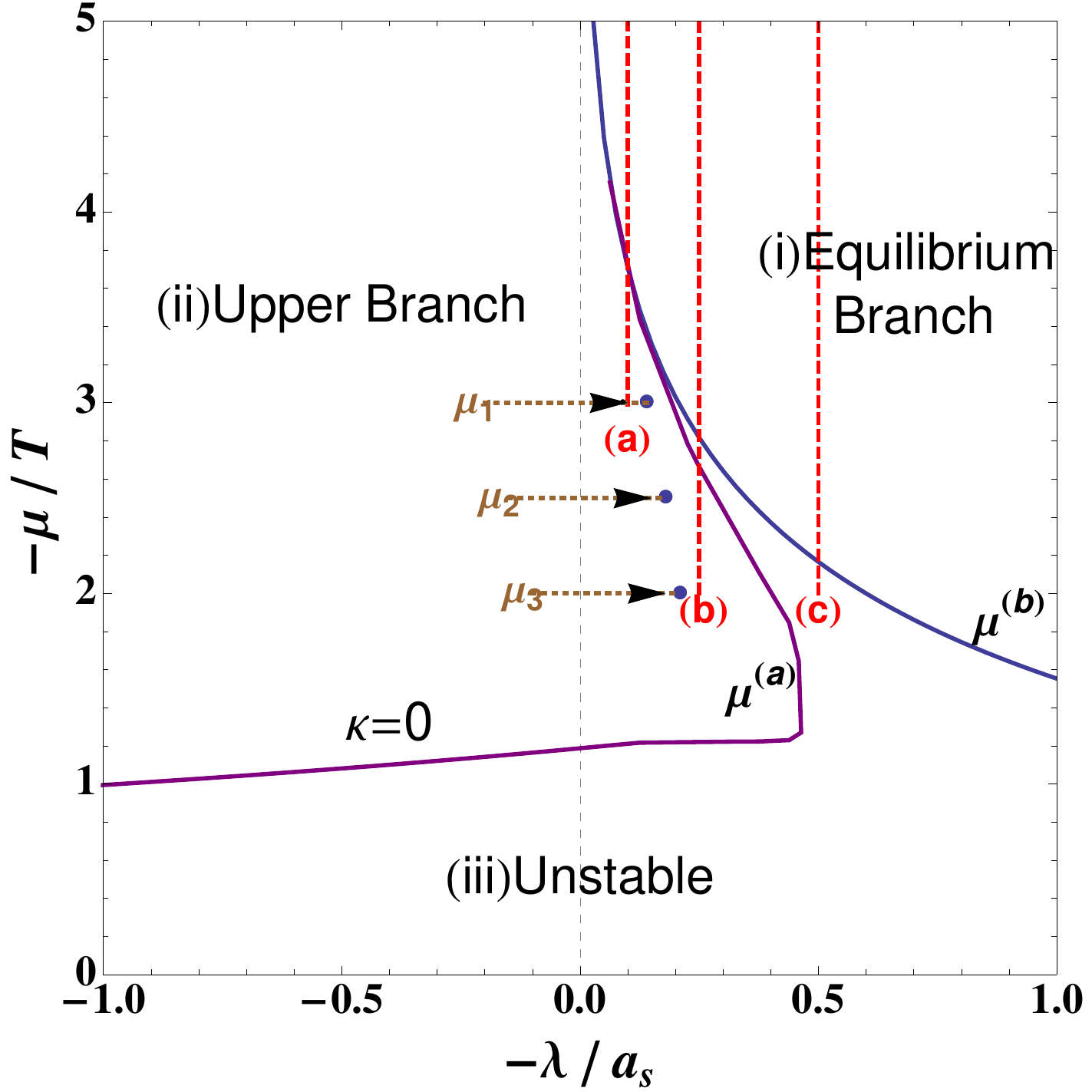}
\caption{\label{fig:trapPD}Phase diagram in a trap at fixed temperature $T$ and trap frequency $\omega$: 
 $\mu^{a}$ represents the state of compressibility $\kappa =0$. $\mu^{(b)}$ separates the equilibrium branch with the unstable region. The latter has $\kappa<0$. 
Each point on this diagram denotes a density profile of the Bose gas, with $\mu$ being the chemical potential at the trap center. The density profile can be generated by an upward vertical trajectory using LDA, (see text). 
The three horizontal lines with arrows are trajectories of a Bose gas across resonance into the atomic side at fixed $\mu$.  The termination point (denoted by a black dot) on each $\mu$-trajectory indicates the critical scattering length $a_{\ast}$ for that $\mu$. For $a_{s}<a_{\ast}$ ($a_{s}>a_{\ast}$), the density profile is stable (unstable).  Hence, the density profiles $(a)$ with chemical $\mu_{1}$ is stable, whereas the densities $(b)$ and $(c)$, with chemical potential $\mu_{3}$ are unstable.  
 For $T=1\mu K$, and $\omega=2\pi (250Hz)$, we find $(-\lambda/a_s)^{\ast}=0.14, 0.18$ and 0.21 for 
the trajectories $-(\mu/T)_1= 3$,  $-(\mu/T)_2=2.5$ and $-(\mu/T)_3=2$ respectively. 
The corresponding  particle numbers are $(N_1, N_2, N_3)= (2.1, 3.4, 5.2)\times 10^4$. All these systems satisfy the condition to be in the low-recombination regime, Eq.(\ref{conditions}) or equivalently Eq.(\ref{Nast}), as $\langle \gamma_{3}\rangle_{ave}/\omega= 0.14, 0.35$, and 0.96 respectively, and $N_1, N_2, N_3 <N^{\ast}$, where 
$N^{\ast}= 6.5 \times 10^{4}$ for this temperature and trap frequency. }
 \end{figure}

{\em (C) Upper branch Bose gas in a trap: } In a trap, the density profile within local density approximation (LDA) is given by $n({\bf r}) = n_{upper}(\mu-V({\bf r}), T)$, where $\mu$ is the chemical potential at the center of the trap.  A global view of the density profile can be obtained from the phase diagram in the $(\mu/T)$-$(\lambda/a_{s})$ plane, 
 Figure \ref{fig:trapPD}, where $\lambda$ is the thermal wavelength. 
The density profile along a radial direction starting from the trap center corresponds to a vertical line emerging from $-{\mu}/T$ upward.  A trapped Bose gas is therefore specified by a point $(-\lambda/a_{s}, -\mu/T)$ on this diagram. 

 Figure \ref{fig:trapPD} describes the behavior of the Bose gas as it is swept from the molecular to the atomic side. Three regions are found  from the equation of state: (i) the equilibrium branch, (ii) the upper branch, and (iii) a region of mechanical instability where  $\kappa<0$. 
 ($\kappa>0$ for both (i) and (ii)).  The boundary between (i) and (iii), and that between (ii) and (iii) will be denoted as $\mu^{(b)}(T)$ and $\mu^{(a)}(T)$ respectively.  $\mu^{(a)}(T)$ is the boundary of zero compressibility, $\kappa=0$.  $\mu^{(b)}(T)$ is the boundary where  bound pairs with zero momentum begin to form. 

\begin{figure}
\includegraphics[width=0.45\textwidth]{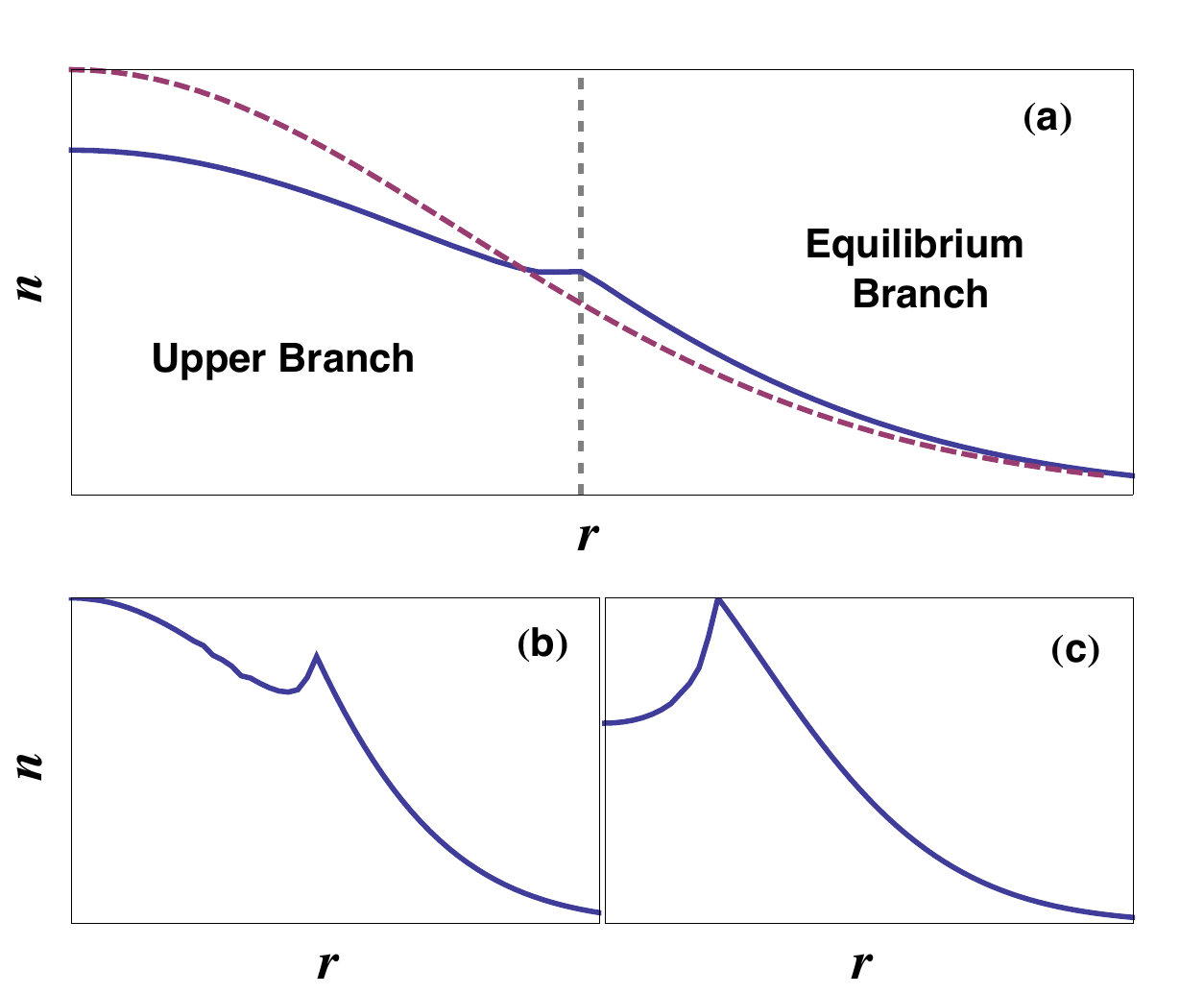}
\caption{\label{fig:profile} The density profiles of an upper branch Bose gas in a trap : 
Fig.(a), (b), and (c) correspond to the densities (a), (b), and (c) in Fig. \ref{fig:trapPD}. Density (a) has an upper branch core and an outer shell of equilibrium state.  The dashed purple curve is the density of the equilibrium branch, Eq.(\ref{n-lower}),   at the same $T$, $a_{s}$, $\omega$, and $N$.  The densities (b) and (c) are unstable as they contain regions where ${\rm d}n/{\rm d}\mu<0$. For density (a), the width of the unstable region becomes less than inter-particle spacing and is therefore non-existent. }
\end{figure}

The unstable region (iii) intervenes between branches (i) and (ii). 
Because of this intervention, any density profile whose center starting from the upper branch on the atomic side will pass through the unstable region, and is therefore unstable.  The width of the unstable region is $\Delta r = r_a - r_b$, where 
$r_a$ and $r_b$ are given by $\mu_a=\mu-V({\bf r}_a)$ and 
$\mu_b=\mu-V({\bf r}_b)$. 
As seen in Fig.\ref{fig:trapPD}, the difference $\mu_b - \mu_a$ (and hence $\Delta r$) becomes very small close to 
resonance.  Should $\Delta r$ be less than inter-particle spacing, i.e. $\Delta r <n(\bar{r})^{-1/3}$, which occurs at a critical ratio $\lambda/a_{s}^{\ast}$ for given $\mu/T$, $r_a$ and $r_b$ can be viewed as a single point $\bar{r}$.  In this case,  the unstable region disappears.  The critical ratio $\lambda/a_{s}^{\ast}$ can be estimated by setting $\Delta r= n(r_a)^{-1/3}$. ($a_{s}^{\ast}$ is a function of $T$ and $\mu$).

Thus, for an upper branch Bose gas characterized by the  point $(-\lambda/a_{s}, -\mu/T)$ on this diagram, it will only be stable when  $-1/a_{s} < -1/a_{s}^{\ast}(T, \mu)$, so that the unstable region with a width $\Delta r$ in real space collapses to zero. The vertical line labelled $(a)$ in Fig.\ref{fig:trapPD} represents such a density.  Its density profile is shown in  Fig.\ref{fig:profile}a, which consists of an upper branch inner core and an equilibrium branch outer layer, both of which are free of Feshbach molecules. Compared to the density profile of the lower branch (dashed line in
Fig.\ref{fig:profile}a which is close to Boltzmann distribution), one sees a discernible kink in the upper branch density. 
When $-\lambda/a_{s}$ exceeds  $-\lambda/a_{s}^{\ast}(T, \mu)$, such as $(b)$ and $(c)$ in Fig, 3, the corresponding density profiles are unstable, for they will contain a region of negative compressibility as shown in  Fig.\ref{fig:profile}b and \ref{fig:profile}c.  
 
From the equation of states for both the upper branch and the equilibrium branch, Eq.(\ref{n-upper}) and (\ref{n-lower}), one can determine the total number of particles once the chemical potential at the center is specified.  
  We also find that total particle number $N$ changes little with $a_{s}$ for given $\mu$ and $T$.   It is straightforward to show that  $N=\int {\rm d}{\bf r}n({\bf r})$ has the general form $N= A(T/\omega)^3$, where $A$ is a  dimensionless number depending on $(-\mu/T, -\lambda/a_s)$. We find that $A<1$. This is expected, as the critical number of Bose-Einstein condensation in a harmonic trap with frequency $\omega$ is $N_{bec}=(0.95)^{-1}(T/\omega)^3$\cite{Pethick}. 
On the other hand, for the trapped gas to be in the low-recombination regime, Eq.(\ref{conditions}) imposes constraints on the central density, and hence the total particle number $N$. To find an estimate of this constraint, we 
approximate the actual density (say that in Fig.4a ) by the Boltzmann form, which then gives  $N\sim 
e^{\mu/T}(T/\omega)^3$\cite{Pethick}. Within the same approximation, we find the quantity $\overline{n}$ in 
Eq.(\ref{conditions}) to be $\overline{n} = 3^{3/4}e^{\mu/T}/\lambda^3$, which then imply 
\begin{equation}
N< N^{\ast} = \alpha  (T/\omega)^{2.5}, \,\,\,\,\,\, \alpha = 3^{3/4}{\cal C}^{-1/2}  = 1.024.
 \label{Nast} \end{equation} 
 We have thus established the results ${\bf (I)}$ to ${\bf (III)}$. 

TLH would like to thank Cheng Chin for discussions of two-body vs three-body collision rates, and thank Nir Navon, Christophe Salomon, Wolfgang Ketterle, and Randy Hulet  for discussions. This work is supported by NSF Grant DMR-0907366 and by DARPA under
the Army Research Office Grant Nos. W911NF-07-1-0464, W911NF0710576,

\end{document}